\documentclass[12pt]{iopart}
\usepackage{amssymb}
\usepackage{rotating}
\usepackage{multirow}
\usepackage{graphicx}
\usepackage{array}
\usepackage{rotfloat}
\usepackage{lscape}

\begin{document}

\title[]{Interplay of symmetries, null forms, Darboux polynomials, integrating factors and Jacobi multipliers in integrable second order differential equations}

\author[Mohanasubha, Chandrasekar, Senthilvelan and Lakshmanan]{R. Mohanasubha, V. K. Chandrasekar, M. Senthilvelan and M. Lakshmanan}

\address{Centre for Nonlinear Dynamics, School of Physics,
Bharathidasan University, Tiruchirappalli - 620 024, India }



\begin{abstract}{Null forms, Symmetries, Darboux polynomials, Integrating factors and Jacobi last multiplier}
In this work, we establish a connection between the  extended Prelle-Singer procedure (Chandrasekar \textit{et al.} Proc. R. Soc. A 2005) with five other analytical methods which are widely used to identify integrable systems in the contemporary literature, especially for second order nonlinear ordinary differential equations (ODEs). By synthesizing these methods we bring out the interplay between Lie point symmetries, $\lambda$-symmetries, adjoint symmetries, null-forms, Darboux polynomials, integrating factors and Jacobi last multiplier in identifying the integrable systems described by second order ODEs. We also give new perspectives to the extended Prelle-Singer procedure developed by us. We illustrate these subtle connections with modified Emden equation as a suitable example. 

\end{abstract}

\section{Introduction}
The last three decades have witnessed a veritable explosion of activities in the theory of integrable systems. Confining our attention only on identifying, classifying and exploring the dynamics of integrable systems, several novel and ingenious methods have been introduced (Babelon \textit{et al.} 2003). Among these, some are reinventions of the integration techniques which were developed in eighteenth and nineteenth centuries by distinguished mathematicians, whereas a few others were introduced to overcome the demerits in some of the earlier ones, and the remaining ones were exclusively developed to meet the contemporary needs. The most versatile and widely used mathematical tools to identify integrable systems belonging to ODEs are (i) Lie symmetry analysis (ii) Darboux polynomials, (iii) Prelle-Singer method, (iv) $\lambda$-symmetries method, (v) adjoint symmetries, (vi) Jacobi last multiplier method and (vii) Painlev$\mathrm{\acute{e}}$ analysis (ARS algorithm). Since the literature is vast we do not recall all the methods here. Even though the methods cited above are apparently different from each other they all essentially seek either one or more of the following aspects, namely symmetries/integrating factors/integrals/solutions (the only exception in the above list is the Painlev$\mathrm{\acute{e}}$ analysis which essentially deals with the singularity structure aspects of the solutions). For example, Lie symmetry analysis, which was originally developed by Sophus Lie in the later part of the nineteenth century, provides an algorithm to determine point symmetries associated with the given equation. Finding integrating factors and integrals from the Lie point symmetries is often a cumbersome procedure. To overcome this difficulty, several generalizations have been proposed. A notable procedure in this direction one can say is the $\lambda$-symmetries method which is applicable when the underlying nonlinear system lacks the required number of Lie point symmetries. The $\lambda$-symmetries method provides a straightforward algorithm to determine more generalized symmetries from which one can proceed to construct integrating factors and integrals for a given second order ODE. The connection between Lie point symmetries and $\lambda$-symmetries has also been elaborated by Muriel and Romero (2009). Recently three of the present authors have developed the extended Prelle-Singer procedure originally introduced by Prelle and Singer for planar differential equations (Prelle \& Singer, 1983) which was extended by Duarte \textit{et al.}, (2001) to second order ODEs. Their approach was based on the conjecture that if an elementary solution exists for the given second-order ODE then there exists at least one elementary first integral $I(t,x,\dot{x})$ whose derivatives are all rational functions of $t$, $x$ and $\dot{x}$. This is applicable to a class of differential equations of any order and any number of coupled ODES (Chandrasekar \textit{et al.} 2005a, 2006, 2009a, 2009b). The method essentially seeks two sets of functions, namely (i) null forms and (ii) integrating factors. An integral (and more integrals) can be obtained from these two sets of functions. On the other hand the method developed by Darboux, now called Darboux polynomials method, provides a strategy to find first integrals (Darboux, 1878). Darboux showed that if we have enough Darboux polynomials then there exists a rational first integral. During the past two decades Llibre and his collaborators (Llibre \& Zhang, 2011; Christopher \textit{et al.} 2007) have studied the Darboux integrability of several nonlinear dynamical systems by exploring Darboux polynomials and their associated integrals, see for example (Ferragut \& Llibre, 2007; Garc$\mathrm{\acute{i}}$a \textit{et al.} 2010) and the references therein.

The Jacobi last multiplier method which was introduced by Jacobi lay dormant for centuries (Jacobi, 1844, 1886) until Nucci and her collaborators have demonstrated the applicability of this method in exploring non-standard Lagrangians associated with certain second order nonlinear ODEs (Nucci \& Leach, 2008). From the known Lie point symmetries of a given equation one can construct a multiplier from which a Lagrangian can be obtained by integration. The adjoint symmetry method which was developed by Bluman and Anco provides an algorithm to determine the integrating factors and their associated integrals of motion (Bluman \& Anco, 2002). This method has also been introduced to overcome some of the demerits of Lie's method.  

As noted above all these methods essentially seek one or more of the following factors, namely symmetries/integrating factors/integrals/solutions. In due course attempts have also been made to interconnect the methods mentioned above. The first attempt in this direction came from Muriel and Romero (2009). They have shown that $\lambda$-symmetries are nothing but the null forms (with a negative sign) given in Prelle-Singer procedure. 
The connection between Lie point symmetries and Jacobi multiplier is reemphasized by Nucci (2005). Similarly the connection between Jacobi multiplier and Darboux polynomials has also been noted by Cari$\mathrm{\tilde{n}}$ena and Ra$\mathrm{\tilde{n}}$ada (2010).

The above discussions clearly show that the interconnections have been made only disconnectedly, for example (i) $\lambda$-symmetries and point symmetries, (ii) multiplier with Darboux polynomials, (iii) $\lambda$-symmetries and null forms and (iv) Lie point symmetries and Jacobi last multiplier. A natural question which arises here is whether there exists a more encompassing interconnection which relates all these methods. Here we answer this broader question in the affirmative.

In this paper we establish a road map between extended Prelle-Singer procedure with all other methods cited above and thereby demonstrate the interplay between Lie point symmetries, $\lambda$-symmetries, adjoint symmetries, null forms, integrating factors, Darboux polynomials, and Jacobi multiplier of integrable systems, at least for the second order ODEs, which can then possibly be extended to higher order ODEs. To achieve this goal we start our investigations with the extended Prelle-Singer procedure. In this procedure, we have basically two equations to integrate. The first one is for the null-form ($S$) and other one is for the integrating factor ($R$). We first interconnect the $S$ equation with Lie point symmetries and $\lambda$-symmetries, by introducing a transformation $S=-\frac{D[X]} {X}$ in the $S$ equation, where $D$ is the total derivative and rewrite the later as a linear second order ODE in $X$. We then show that this second order equation is nothing but the Lie's invariance condition for the given second order ODE in terms of the characteristic vector field $Q=\eta-\dot{x}\xi$ with $Q=X$. Thus solving this linear second order ODE one not only gets Lie point symmetries $\xi$ and $\eta$ along with the characteristics but also the null form $S$. We then introduce another transformation $R=\frac{X} {F}$, and rewrite the $R$ equation in a new variable $F$. We then show that this function $F$ is nothing but a function equivalent to the Darboux polynomials. Through this relation we are able to establish a direct connection between integrating factors which arise in the extended Prelle-Singer procedure with Darboux polynomials. 
 This in turn connects the integrating factors with Jacobi multiplier as well. This remarkable connection, $R=\frac{X} {F}$, namely integrating factors are quotient functions in which the numerator is connected to Lie point symmetries/$\lambda$-symmetries/null forms and the denominator is connected to Darboux polynomials/Jacobi last multiplier brings out the hidden connection between the quantities that determine the integrability. Finally by rewriting the $S$ and $R$ equations, vide Eqs.(\ref{met9}) and (\ref{met10}), given below as a single second order ODE the latter becomes the determining equation for adjoint symmetry equation. This in turn confirms that the adjoint symmetries are nothing but the integrating factors for this restricted class of ODEs, that is adjoint symmetries should satisfy the adjoint invariance conditions. In this case adjoint symmetries are nothing but the integrating factors. By establishing these conditions we bring out the interplay between the various quantitative factors determining integrability.

The plan of the paper is as follows.
In section 2, we describe the Prelle-Singer procedure for solving second order differential equations. In addition, we demonstrate the connection between the several well known methods such as $\lambda-$symmetries, Darboux polynomials, Jacobi last multiplier and adjoint symmetries methods. In section 3, we prove the relationship between the several methods with the PS method with an example, namely modified Emden equation (MEE). Finally we summarize our results in section 4. 
\section{Extended Prelle-Singer method for second order ODEs}
\label{sec2}
In this section, we briefly discuss the modified Prelle-Singer procedure for second order ODEs (Chandrasekar V K \textit{et al.} 2005a; Duarte \textit{et al.}, 2001).  Let us 
consider the second order ODEs of the form 
\begin{equation} 
\ddot{x}=\frac{P}{Q}=\phi,  \quad {P,Q}\in \mathbb{C}{[t,x,\dot{x}]}, \label{met1}
\end{equation}
where over dot denotes differentiation with respect to time and $P$ and $Q$ 
are analytic functions of the variables $t$, $x$ and $\dot{x}$.  Let us assume that the ODE (\ref{met1}) 
admits a first integral $I(t,x,\dot{x})=C,$ with $C$ constant on the 
solutions, so that the total differential gives
\begin{eqnarray}  
dI={I_t}{dt}+{I_{x}}{dx}+{I_{\dot{x}}{d\dot{x}}}=0, 
\label{met3}  
\end{eqnarray}
where the subscript denotes partial differentiation with respect 
to that variable.  Rewriting equation~(\ref{met1}) in the form 
$\frac{P}{Q}dt-d\dot{x}=0$ and adding a null term 
$S(t,x,\dot{x})\dot{x}$ $ dt - S(t,x,\dot{x})dx$ to the latter, we obtain that on 
the solutions the 1-form
\begin{eqnarray}
\bigg(\frac{P}{Q}+S\dot{x}\bigg) dt-Sdx-d\dot{x} = 0. 
\label{met6} 
\end{eqnarray}	
Hence, on the solutions, the 1-forms (\ref{met3}) and 
(\ref{met6}) must be proportional.  Multiplying (\ref{met6}) by the 
factor $ R(t,x,\dot{x})$ which acts as the integrating factor
for (\ref{met6}), we have on the solutions that 
\begin{eqnarray} 
dI=R(\phi+S\dot{x})dt-RSdx-Rd\dot{x}=0, 
\label{met7}
\end{eqnarray}
where $ \phi\equiv {P}/{Q}$.  Comparing equations (\ref{met3}) 
with (\ref{met7}) we end up with the three relations which relates the integral($I$), integrating factor($R$) and the null term($S$). 
\begin{eqnarray} 
I_{t} &=& R(\phi+\dot{x}S), \nonumber\\
I_{x} &=& -RS, \nonumber\\
I_{\dot{x}} &=& -R.  
\label{met8}
\end{eqnarray} 
Then the compatibility conditions between these variables gives the determining equations to find $R$ and $S$ which are given in the following equations. 
\begin{eqnarray}  
D[S]=&-\phi_x+S\phi_{\dot{x}}+S^2,\label{met9}\\
D[R]=&-R(S+\phi_{\dot{x}}),\label{met10}\\
R_x=&R_{\dot{x}}S+RS_{\dot{x}},\label{met11}
\end{eqnarray}
where 
\begin{eqnarray}
D=\frac{\partial}{\partial{t}}+
\dot{x}\frac{\partial}{\partial{x}}+\phi\frac{\partial}
{\partial{\dot{x}}}.
\nonumber
\end{eqnarray}

Equations~(\ref{met9})-(\ref{met11}) can be solved in principle in the following way.  
From (\ref{met9}) we can find $S$. Once $S$ is known then  
equation~(\ref{met10}) becomes the determining equation for the function $R$.  
Solving the latter one can get an explicit form for $R$.  
Now the functions $R$ and $S$ have to satisfy an extra constraint, that is, 
equation~(\ref{met11}).  Once a compatible solution satisfying all the three 
equations have 
been found then the functions $R$ and $S$ fix the integral of motion 
$I(t,x,\dot{x})$ by the relation 
\begin{eqnarray}\nonumber
I(t,x,\dot{x})= \int R(\phi+\dot{x}S)dt
 -\int \left( RS+\frac{d}{dx}\int R(\phi+\dot{x}S)dt \right) dx\\ \nonumber
 -\int \left\{R+\frac{d}{d\dot{x}} \left[\int R (\phi+\dot{x}S)dt-
\int \left(RS+\frac{d}{dx}\int R(\phi+\dot{x}S)dt\right)dx\right] 
\right\}d\dot{x}.\\ \label{met13}
\end{eqnarray}
Equation~(\ref{met13}) can be derived straightforwardly by 
integrating the three relations which relates the integral($I$), integrating factor($R$) and the null term($S$).
Note that for every independent set $(S,R)$, equation~(\ref{met13}) defines an integral.

\subsection{Interconnections}
To demonstrate that the null form $S$ and the integrating factor $R$ are intimately related with other measures of integrability we do the following. 
By introducing a transformation 
\begin{equation}
S=-D[X]/X\label{sx},
\end{equation} Eq.(\ref{met9}) becomes a linear equation in the new variable $X$, that is
\begin{equation}  
D^2[X]  = \phi_{\dot{x}} D[X]+\phi_x X,
\label{met14}
\end{equation}
where $D$ is the total differential operator. With another change of variable
\begin{equation}
R=X/F,\label{rxf}
\end{equation}
where $F(t,x,\dot{x})$ is a function to be determined, we can rewrite Eq.(\ref{met10}) in a compact form in the new variable $F$ as
\begin{equation}  
D[F] = \phi_{\dot{x}}F.
\label{met15}
\end{equation}
In the following, we demonstrate that the functions $X$ and $F$ are intimately related to Lie symmetries, $\lambda$-symmetries, Darboux polynomials, Jacobi last multiplier and adjoint symmetries.


\subsection{Connection between Lie Symmetries and null forms}
Let $v=\xi\partial_t+\eta\partial_x$ be the Lie point symmetry generator of (\ref{met1}), when $\xi(t,x)$ and $\eta(t,x)$ are infinitesimals associated with the $t$ and $x$ variables, respectively. Then the characteristic of $v$ is given by $Q=\eta-\dot{x}\xi$ (Olver 1995). Then the infinitesimal operator associated with the vector field $v$ is $v^{(2)}=\xi\partial_t+\eta\partial_x
+\eta^{(1)}\partial_{\dot{x}}+\eta^{(2)}\partial_{\ddot{x}}$, where $\eta^{(1)}$ and $\eta^{(2)}$ are the first and second prolongations of the vector field $v$ and are given by $\eta^{(1)}=\dot{\eta}-\dot{x}\dot{\xi}$ and $\eta^{(2)}=\ddot{\eta}-\ddot{x}\dot{\xi}$, where over dot denotes total differentiation with respect to $t$.

The invariance condition of the second order ODE, $\ddot{x}=\phi(t,x,\dot{x})$, determines the infinitesimal symmetries $\xi$ and $\eta$ explicitly, through the condition $v^{(2)}[\ddot{x}-\phi(t,x,\dot{x})]=0$ (Olver 1995). Expanding the later, one finds the invariance condition in terms of the above evolutionary vector field $Q$ as
\begin{equation}  
D^2[Q]  = \phi_{\dot{x}} D[Q]+\phi_x Q.
\label{met16}
\end{equation}
Comparing Eqs.(\ref{met14}) and (\ref{met16}) we find that
\begin{equation}
X=Q.\label{xq}
\end{equation} 
In other words the $S$-determining equation (\ref{met14}) now becomes exactly the determining equation for the Lie point symmetries $\xi$ and $\eta$ with $Q=\eta-\dot{x}\xi$. Since $S=-\frac{D[X]} {X}$, the null form $S$ can also be determined once $\xi$ and $\eta$ are known. This establishes the connection between the null forms $S$ with the Lie symmetries ($\xi$ and $\eta$).

\subsection{Connection between $\lambda$- symmetries and null forms}
All the nonlinear ODEs do not necessarily admit Lie point symmetries.  Under such a circumstance one may look for  generalized symmetries associated with the given equation.  One such generalized symmetry is the $\lambda$-symmetry.  The $\lambda$-symmetries can be derived by a well defined algorithm which includes Lie point symmetries as a very specific subclass and have an associated order reduction procedure which is similar to the classical Lie method of reduction.  Although $\lambda$-symmetries are not Lie point symmetries, the unique prolongation of vector fields  to the space of variables $(t,x,...,x^n)$ for which Lie reduction method applies is always a $\lambda$-prolongation for some function $\lambda(t,x,\dot x)$. For more details, one may see the works of Muriel and Romero (Muriel \& Romero, 2001, 2008, 2009, 2012).\\
Now, if we replace $S=-Y$ in equation (\ref{met9}) we get
\begin{equation}  
D[Y]  = \phi_x+Y\phi_{\dot{x}}-Y^2,\label{met17}
\end{equation}
which is nothing but the determining equation for the $\lambda$-symmetries for a second order ODE (Muriel \& Romero 2009) which in turn establishes the connection between $\lambda$-symmetries and null forms. It has also been shown (Muriel \& Romero 2009) that once Lie point symmetries are known, the $\lambda$-symmetries can be constructed through the relation $Y=D[Q]/Q$ which is also confirmed here.

\subsection{Connection between Darboux polynomials and integrating factors}
Here we recall briefly the role of Darboux polynomials. Let us consider the function $G=\prod f_i^{n_i}$, where $f_i'$s are Darboux polynomials and $n_i'$s are rational numbers. If we can identify a sufficient number of Darboux polynomials (irreducible polynomials) $f_i'$s, satisfying the relations $D[f_i]/f_i=\alpha_i$, where $\alpha_i$'s are the co-factors, then 
\begin{equation}
D[G]/G=\sum_in_i\frac{D[f_i]}{f_i}=n_i\alpha_i.\label{laseq}
\end{equation} Suppose $f_1$ and $f_2$ are two Darboux polynomials for (\ref{met1}) with the same cofactor $\alpha_i$, then the ratio $f_1/f_2$ defines an integral.

Now we compare Eq.(\ref{laseq}) with Eq.(\ref{met15}). If we choose $G=F$ then equation (\ref{met15}) becomes
\begin{equation}  
\sum_in_i\frac{D[f_i]}{f_i}=\phi_{\dot{x}}.\label{met22}
\end{equation}
In other words the Darboux polynomials constitute the solution of Eq.(\ref{met15}). Since $X$ is already known, once Darboux polynomials are known the integrating factors can be fixed (vide Eq.(\ref{rxf})). 

\subsection{Connection between Jacobi last multiplier and Darboux polynomials/Integrating factors}
Let us rewrite the second order ODE (\ref{met1}) into an equivalent system of two first-order ODEs $\dot{x}_i=w_i(x_1,x_2)$, $i=1,2$. Then its Jacobi last multiplier $M$ is obtained by solving the following differential equation (Nucci 2005),
\begin{equation}  
\frac{\partial M}{\partial t}+\sum_{i=1}^2\frac{\partial (Mw_i)}{\partial x_i}=0\label{met18}.
\end{equation}
The above equation can be rewritten as
\begin{equation}  
D[\log M]+\sum_{i=1}^2\frac{\partial w_i}{\partial x_i}=0.\label{met19}
\end{equation}
For the present case (\ref{met1}), Eq.(\ref{met19}) is further simplified to
\begin{equation}  
D[\log M]+\phi_{\dot{x}}=0.\label{met20}
\end{equation}
Now comparing the equation (\ref{met15}) with (\ref{met20}) we find that $F=M^{-1}$. Thus from the knowledge of the multiplier $M=\big(\frac{1} {\Delta}\big)$ we can also fix the explicit form of $F$ which appears in the denominator of integrating factor $R$ (vide Eq.(\ref{rxf})) as $F=\Delta$ or vice versa.

It has also been shown that the multiplier for the second order ODE is determined from (Nucci 2005)
\[ \Delta = \left| \begin{array}{ccc}
1 & \dot{x} & \ddot{x} \\
\xi_1 & \eta_1 & \eta_{1}^{(1)} \\
\xi_2 & \eta_2 & \eta_{2}^{(1)} \end{array} \right|.\]\label{delta}

where $(\xi_1,\eta_1)$ and $(\xi_2,\eta_2)$  are two sets of Lie point symmetries of the second order ODE and $\eta_{1}^{(1)}$ and $\eta_{2}^{(1)}$ are the corresponding prolongations. The determinant establishes the connection between the multiplier and Lie point symmetries. Once the multiplier is known its inverse provides the function $F$ which in turn forms the denominator of the integrating factor. 

We also recall here that $M$ is related to the Lagrangian $L$ through the relation (Jacobi 1886)
\begin{equation}  
M=\frac{\partial^2 L}{\partial \dot{x}^2}=\frac{1} {F}.\label{met21}
\end{equation}
With the known expression of $M$ or $F$, the Lagrangian $L$ can be obtained by straightforward integration.

\subsection{Connection between adjoint symmetries and integrating factors}
Considering the second order ODE (\ref{met1}),~the linearized symmetry condition (vide Eq.(\ref{met16})) is given by
\begin{equation}
L[x]v=D^2[v]-\phi_{\dot{x}}D[v]-\phi_{x}v=0,
\end{equation}
and the corresponding adjoint of the linearized symmetry condition is given by
\begin{equation}
L^*[x]w=D^2[\Lambda]+D[\phi_{\dot{x}}\Lambda]-\phi_{x}\Lambda=0.\label{met23}
\end{equation}
Let us rewrite the coupled equations (\ref{met9}) and (\ref{met10}) into an equation for the single function $R$. Then the resultant equation turns out to be of the form
\begin{equation}  
D^2[R]+D[\phi_{\dot{x}}R]-\phi_{x}R=0.\label{met24}
\end{equation}

Comparing the above two equations (\ref{met23}) and (\ref{met24}) one can conclude that the integrating factor $R$ is nothing but the adjoint symmetry $\Lambda$, that is 
\begin{equation}
R=\Lambda. \label{met25}
\end{equation}
Thus the integrating factor turns out to be the adjoint symmetry of the given second order ODE.

In the following section, we demonstrate the above connections with a suitable example.
\section{Example}
Let us consider a model which is of contemporary interest in integrable systems, namely the modified Emden equation (Mahomed \& Leach 1985; Leach \textit{et al.} 1988; Chandrasekar \textit{et al.}, 2005a, 2005b, 2007; Cari$\mathrm{\tilde{n}}$ena \textit{et al.} 2005, 2010; Nucci 2012). The equation of motion is given by
\begin{equation}
\ddot{x}=-3x\dot{x}-x^3.\label{exam}
\end{equation}
This equation arises in the study of equilibrium configurations of a spherical gas cloud acting under the mutual attraction of its molecules and subject to the laws of thermodynamics and in the modeling of the fusion of pellets. This system also admits time independent nonstandard Lagrangian and Hamiltonian functions. For a more general equation, see Chandrasekar et al. (2005b). Eq.(\ref{exam}) is also known as the Riccati second-order equation in the literature (see for example, W. F. Ames 1968).  \\
Now, for Eq.(\ref{exam}), the defining equations (\ref{met9})-(\ref{met11}) for the null form $S$ and integrating factor $R$ become
\begin{eqnarray}
S_t+\dot{x}S_x-(3x\dot{x}+x^3)S_{\dot{x}}&=&(3\dot{x}+3x^2)-3S_x+S^2,\label{n1} \\ 
R_t+\dot{x}R_x-(3x\dot{x}+x^3)R_{\dot{x}}&=&-R(S-3x),\label{n2} \\ 
R_x&=&R_{\dot{x}}S+RS_{\dot{x}}. \label{n3}
\end{eqnarray}
Two sets of explicit forms have been given by Chandrasekar \textit{et al.} (2005a) for the functions $S$ and $R$ as 
\begin{equation}
S_1 =\frac{-\dot{x}+x^2}{x}, \quad R_1= \frac{x} {(\dot{x}+x^2)^2}
\label{lam107}
\end{equation}
and
\begin{equation}
S_2=\frac{2-\dot{x}t^2-4tx+t^2x^2} {t(-2+tx)},~~R_2=\frac{t(-2+tx)} {2(t\dot{x}-x+tx^2)^2}.
\end{equation}
We can also find the above forms of $S$ and $R$ by using the above discussed interconnections. 
\subsection{Lie point symmetries and characteristics}
As we noted earlier in Sec.2.1, we try to solve Eq.(\ref{met14}) which is equivalent to solving Eq.(\ref{n1}). So we start our analysis by solving the same Eq.(\ref{met14}) but now for the characteristics $Q=\eta-\dot{x}\xi$, using the relation $Q=X$.
Substituting this in (\ref{met16}) and equating the various powers of $\dot{x}$, we get a set of partial differential equations for $\xi$ and $\eta$. Solving them consistently we find explicit expressions for $\xi$ and $\eta$. In our case Eq.(\ref{exam}) admits eight dimensional Lie point symmetries. The corresponding vector fields are: (see also Leach \textit{et al.} 1988; Pandey \textit{et al.} 2009)
\begin{equation}
V_1=x\frac{\partial} {\partial{t}}-x^3\frac{\partial} {\partial{x}},\nonumber
\end{equation}
\begin{equation}
V_2=t^2\bigg(1-\frac{xt} {2}\bigg)\frac{\partial} {\partial{t}}+xt\bigg(1-\frac{3} {2}xt+\frac{x^2 t^2} {2}\bigg)\frac{\partial} {\partial{x}},\nonumber
\end{equation}
\begin{equation}
V_3=\frac{\partial} {\partial{t}},~~~~V_4=t\bigg(1-\frac{xt} {2}\bigg)\frac{\partial} {\partial{t}}+x^2t\bigg(-1+\frac{xt} {2}\bigg)\frac{\partial} {\partial{x}},\nonumber
\end{equation}
\begin{equation}
~~~V_5=xt\frac{\partial} {\partial{t}}+x^2\bigg(1-xt\bigg)\frac{\partial} {\partial{x}},\nonumber
\end{equation}
\begin{equation}
V_6=-\frac{xt^2} {2}\frac{\partial} {\partial{t}}+x\bigg(1-xt+\frac{x^2 t^2} {2}\bigg)\frac{\partial} {\partial{x}},\nonumber
\end{equation}
\begin{equation}
V_7=\frac{3} {2}t^2\bigg(1-\frac{xt} {3}\bigg)\frac{\partial} {\partial{t}}+\bigg(1-\frac{3} {2}x^2t^2+\frac{x^3 t^3} {2}\bigg)\frac{\partial} {\partial{x}},\nonumber
\end{equation}
\begin{equation}
V_8=-\frac{t^3} {2}\bigg(1-\frac{xt} {2}\bigg)\frac{\partial} {\partial{t}}+t\bigg(1-\frac{3} {2}xt+x^2t^2-\frac{x^3t^3} {4}\bigg)\frac{\partial} {\partial{x}}.\label{vec1}
\end{equation}
Since we are dealing with a second order ODE, we consider any two vector fields to generate all other factors. We consider the vector fields $V_1$ and $V_2$ in the following. The results which arise from other pairs of symmetry generators are summarized in Tables $1$.
\subsection{Lie symmetries and null forms}
From the vector fields $V_1$ and $V_2$ one can identify two sets of infinitesimals $\xi$ and $\eta$ as
\begin{equation}
\xi_1=x,\;\;\eta_1=-x^3,\label{lie1}
\end{equation}
\begin{equation}
\xi_2=t^2\bigg(1-\frac{xt}{2}\bigg),\;\;\eta_2=xt\bigg(1-\frac{3xt} {2}+\frac{x^2t^2}{2}\bigg).\label{lie2}
\end{equation}
The associated characteristics $Q_i=\eta_i-\dot{x}\xi_i$, $i=1,2$, are found to be 
\begin{equation}
Q_1=-\dot{x}x- x^3,\;\;\;\;Q_2=\frac{1} {2}t(-2+xt)(\dot{x}t-x+tx^2).\label{q1q2}
\end{equation}
Recalling the relation $Q=X$ and $S=-D[X]/X$ the null forms $S_1$ ans $S_2$ can be readily found. Our analysis shows that
\begin{equation}
S_1=x-\frac{\dot{x}} {x},\;\;\;\;\;\;\;S_2=\frac{(2-\dot{x}t^2-4tx+t^2x^2)} {t(-2+tx)}.\label{s12}
\end{equation}
One can easily check that $S_1$ and $S_2$ are two particular solutions of (\ref{n1}).
\subsection{$\lambda$- Symmetries and null forms}
Using the relation $S=-\lambda$ we find
\begin{equation}
\lambda_1=-S_1=\frac{\dot{x}} {x}-x,\;\;\;\;\;\;\; \lambda_2=-S_2=-\frac{(2-\dot{x}t^2-4tx+t^2x^2)} {t(-2+tx)}.
\end{equation}
Again it is a straightforward exercise to check that the $\lambda_i'$s, $i=1,2$, indeed satisfy Eq.(\ref{met17}).
\subsection{Darboux polynomials and integrating factors}
As we noted earlier the integrating factors can be derived in two different ways, namely either by exploring Darboux polynomials or by constructing the last multiplier. We consider both the possibilities and demonstrate that both of them lead to the same results.

First we derive the integrating factors from the Darboux polynomials. 
It is straightforward to check that $f_1=-(x^2+\dot{x})$, $f_2=-x+tx^2+t\dot{x}$ and $f_3=1+\frac{\dot{x}t^2} {2}-t x+\frac{t^2x^2} {2}$ are Darboux polynomials of (\ref{exam}) with the same cofactors $\alpha_1=\alpha_2=\alpha_3=-x$. We note here that Eq.(\ref{exam}) also admits two more Darboux polynomials, $f_4=-2 \dot{x}+\dot{x}^2t^2-2 \dot{x}t x+2 \dot{x}t^2 x^2-2tx^3+t^2x^4$ and $f_5=-6 \dot{x}+\dot{x}^2t^2-2 \dot{x}tx-2x^2+2\dot{x}t^2x^2-2tx^3+t^2x^4$ with the same cofactors $-2x$. It is known that combinations of the Darboux polynomials are also Darboux polynomials (Dumortier \textit{et al.} 2006). For illustrative purpose let us consider the polynomials $f_1$ and $f_2$. From Eq.(\ref{met22}) we can evaluate different combinations of $n_i'$s. Using Eq.(\ref{met22}) we find
\begin{equation}
n_1(-x)+n_2(-x)=-3x.\label{n11}
\end{equation}
There are four possible combinations of $n_1$ and $n_2$, namely (3,0), (0,3), (2,1) and 1,2), fulfill the condition (\ref{n11}). Since $F=f_{1}^{n_1}f_{2}^{n_2}$ we obtain four different forms of $F$. The corresponding $F_i'$s, $i=1, 2, 3,4,$ are given by
\begin{eqnarray}
F_1&=&-(\dot{x}+x^2)^3,\label{f3}\\
F_2&=&(-x+tx^2+t\dot{x})^3,\label{fj}\\
F_3&=&(x^2+\dot{x})^2(-x+tx^2+t\dot{x}),\label{f1}\\
F_4&=&-(x^2+\dot{x})(-x+tx^2+t\dot{x})^2.\label{f2}
\end{eqnarray}
From $F_1$ and $F_2$ we construct $R_1$ and $R_2$. The result turns out to be
\begin{equation}
R_1=\frac{x} {(x^2+\dot{x})^2},\;\;\;R_2=\frac{-t(1-\frac{tx} {2})} {(t\dot{x}-x+tx^2)^2}.\label{r22}
\end{equation}
We mention here that the denominator of $S_1$ is the same as the numerator of $R_1$ and also the numerator of $R_2$ matches with the denominator of $S_2$. Since $F_1$ and $F_2$ are two simple Darboux polynomials admitted by Eq.(\ref{exam}) we stick to these polynomials. The role of other Darboux polynomials in determining integrating factors will be presented in the form of tabulation. We discuss the role of them in Table $2$. 

As we noted earlier, the Darboux polynomials can also be derived from the Jacobi last multiplier. Since we already derived Lie point symmetries of (\ref{exam}), we can exploit the connection between Lie point symmetries and Jacobi last multiplier $M$ to deduce the Darboux polynomials. For this purpose let us evaluate the multiplier $M$ which is given by $M=\Delta^{-1}$, provided that $\Delta\neq0$, where $\Delta$ is given by the expression (\ref{delta}).
Since we need two Lie symmetries to evaluate Jacobi last multiplier, see Eq.(\ref{delta}), we choose the vector fields $V_3$ and $V_1$, to obtain first Jacobi last multiplier $M_1$. One can choose the other vector fields also but the determinant should be non zero. Evaluating the associated determinant with these two vector fields, we find
\[ \Delta = \left| \begin{array}{ccc}
1 & \dot{x} & -3x\dot{x}-x^3 \\
x & -x^3  & -\dot{x}(\dot{x}+3x^2)\\
1 & 0 & 0 \end{array} \right|=-(x^2+\dot{x})^3,\]

from which we obtain
\begin{equation}
M_1=-\frac{1} {(x^2+\dot{x})^3}\label{mmm1}.
\end{equation}
To determine $M_2$, we choose the vector fields $V_6$ and $V_4$ which in turn provide $M_2$ in the form
\begin{equation}
M_2=\frac{1} {(-x+tx^2+t\dot{x})^3}\label{mmm2}.
\end{equation}
Now exploiting the relation $F=M^{-1}$ we can obtain the exact forms of $F_1$ and $F_2$  which in turn exactly matches with the one given in Eqs.(\ref{f3}) and (\ref{fj}). \\Since we know the multipliers, we can also construct the associated Lagrangians by straightforward integration by recalling the expressions $M_1=\frac{\partial^2 L_1}{\partial \dot{x}^2}$ and $M_2=\frac{\partial^2 L_2}{\partial \dot{x}^2}$. The resultant Lagrangians are found to be (Nucci \& Tamizhmani, 2010)
\begin{eqnarray}
L_1&=&-\frac{1} {2(\dot{x}+x^2)}+\dot{g}(t,x),\\
L_2&=&\frac{1} {2 t^2 (\dot{x} t + x (-1 + t x))}+\dot{g}(t,x),
\end{eqnarray}
where $g(t,x)$ is the gauge function and dot stands for the total time derivative (Nucci \& Leach, 2008a).
\subsection{Connection between adjoint symmetries and integrating factors}
Finally, we present the adjoint symmetries of (\ref{exam}) which are nothing but the integrating factors of the given equation, namely 
\begin{equation}
\Lambda_1=R_1=\frac{x} {(x^2+\dot{x})^2},\;\;\;\;\ \Lambda_2=R_2=\frac{t(-1+\frac{tx} {2})} {(t\dot{x}-x+tx^2)^2}.
\end{equation}
For the sake of completeness we also present the integrals of Eq.(\ref{exam}). To construct them we use the expression (\ref{met13}). By plugging (\ref{s12}) and (\ref{r22}) in (\ref{met13}) and evaluating the integrals we arrive at the following two integrals, that is
\begin{equation}
I_1=-t+\frac{x} {x^2+\dot{x}},\;\;\;\;I_2=-\frac{t} {2}+\frac{(-1+\frac{tx} {2})} {t\dot{x}-x+tx^2}. \label{int12}
\end{equation} 
As we noted earlier, whenever the Darboux polynomials share the same cofactors then their ratio defines a first integral. The integral $I_1$  given above comes from the ratio of the Darboux polynomials $f_1$ and $f_2$. The ratio of two Jacobi last multipliers also constitute a first integral. For example the first integral $I$ which comes out from the ratio of the multipliers, $(\ref{mmm1})$ and $(\ref{mmm2})$, matches with the integral $I_1$. From the integrals (vide Eq.(\ref{int12})), we can derive the general solution of (\ref{exam}) as
\begin{equation}
x(t) =\frac{(t + I_1)} {(\frac{t^2} {2}+tI_1+I_1I_2)}.
\end{equation}
\subsection{Interconnection between various quantities}
In this sub-section, we summarize the results given in Tables $1-3$ where we have given the null forms, characteristics, vector fields, integrating factors, Darboux polynomials and first integrals of (\ref{exam}). As we have pointed out in the introduction, once we know the quantities $S$ and $R$, one can find all the other quantities using the relations given in Eqs.(\ref{sx}) and (\ref{rxf}) respectively. For example to obtain the expression for $X$ from $S$ one has to integrate the first order partial differential equation (\ref{sx}). As far as the present example is concerned we make an ansatz for $X$ of the form
\begin{equation}
X=a(t,x)\dot{x}+b(t,x).\label{anz}
\end{equation}
Substituting this ansatz in Eq.(\ref{sx}) and equating it to $S_1$ and solving the resultant equations, we get the following characteristics $X$,  that is
\begin{eqnarray}
X_1&=&-x(\dot{x}+x^2),\nonumber\\
X_2&= &-x(\dot{x}t-x+tx^2),\nonumber\\
X_3&=& \frac{1} {2}x(2+\dot{x}t^2-2tx+t^2x^2).
\end{eqnarray} 
Since $X_i=\eta_i-\dot{x}\xi_i$ one can straightforwardly identify the infinitesimal generators/vector fields. We find that the above characteristics correspond to the vector fields $V_1$, $V_5$ and $V_6$ given in Eq.(\ref{vec1}). On the other hand substituting the ansatz (\ref{anz}) in Eq.(\ref{sx}) and equating it with $S_2$, after some algebra we find that they lead to the following characteristics, namely 
\begin{eqnarray}
X_4&=&\frac{1} {2}t(-2+tx)(\dot{x}+x^2),\nonumber \\
X_5&=&\frac{1} {2}t(-2+tx)(\dot{x}t-x+tx^2), \nonumber\\
X_6&=&-\frac{t} {4}(-2+tx)(2+\dot{x}t^2-2tx+t^2x^2),
\end{eqnarray}
which correspond to the vector fields $V_2, V_4$ and $V_8$ respectively. To capture the remaining vector fields/characteristics we consider the other forms of $S$, namely $S_3$ and $S_4$ (which are given in Table $1$) and the corresponding characteristics. Repeating the analysis, we obtain 
\begin{eqnarray}
X_7&=&-\dot{x},\\
X_8&=&\frac{1} {2}(2-3t^2x^2+t^3x^3+\dot{x}t^2(-3+tx)).
\end{eqnarray}
The characteristics are presented in the third column of Table $1$. From $R$ and $X$ one can derive the Darboux polynomials using the relations (\ref{met15}) and (\ref{met22}). One can also derive the Darboux polynomials from Jacobi last multiplier which is given in the fifth column in Table 1. Eq.(\ref{met21}) shows that the Jacobi last multiplier is nothing but the inverse of the Darboux polynomials. One can also derive Jacobi last multiplier from the Lie point symmetries by using the relation (\ref{delta}). Once Jacobi last multiplier is known, we can find the corresponding Lagrangian by Eq.(\ref{met21}). Further, the integrating factor is nothing but the adjoint symmetry as seen from Eq.(\ref{met25}) and the $\lambda$-symmetries are same as the null forms with negative sign.

In Table $2$, we present the complete role of Darboux polynomials in determining the integrating factors. To illustrate this let us consider all the combinations of $f_1$ and $f_2$ alone (vide Eqs.(\ref{f3})-(\ref{f2})). In other words, we only consider $F_i's, i=1,2,3,4$ (vide Eqs.(\ref{f3})-(\ref{f2})) associated with the Darboux polynomials $f_1$ and $f_2$. The integrating factors which come out from these Darboux polynomials are denoted as $R_{ij},i=1,2...8$ and $j=1,2,3,4$. In $R_{ij}$ the first subscript $(i)$ denotes the vector field and the second subscript $(j)$ denotes the Darboux polynomials. All these integrating factors satisfy the first two conditions, that is (\ref{met9}) and (\ref{met10}), in the Prelle-Singer procedure. Suppose the integrating factor and the corresponding null form also satisfy the third equation (\ref{met11}), one can proceed to derive the integral straightforwardly from Eq.(\ref{met13}). For example, the null form $S$, with each one of the integrating factors $R_{11}, R_{12}, R_{13}$ and $R_{14}$ separately satisfy the Eq.(\ref{met11}). The compatible sets $(S_1, R_{11}),(S_1, R_{12}), (S_1, R_{13})$ and $(S_1, R_{14})$ straightforwardly yield the integrals $I_{11}, I_{12}, I_{13}$ and $I_{14}$.

One may observe that some of the integrating factors do not satisfy the third constraint (\ref{met11}). In those cases one can use the first integral derived from the set $(S_1, R_1)$ to deduce a compatible solution (for more details one may refer to Chandrasekar \textit{et al.} (2005a)). 
The integrating factors which come out from this category are denoted as $\hat{R}$ (in order to be consistent with our earlier work). This $\hat{R}$ combined with the null form $S$ satisfies all the three equations (\ref{met9})-(\ref{met11}) in the Prelle-Singer procedure. For example, let us consider the null form associated with vector field $V_2$. This null form when combined with $R_{22}$ satisfies the third equation (\ref{met11}) straightforwardly. The other three integrating factors $R_{21}, R_{23}$ and $R_{24}$ coming out from (\ref{met10}) do not satisfy the third equation (\ref{met11}). For these three cases we have followed the above said procedure and determine the suitable integrating factors that also satisfy the Eq.(\ref{met11}). The compatible integrating factors are denoted as $\hat{R}_{21}, \hat{R}_{23}$ and $\hat{R}_{24}$. Similar arguments are also followed for all the other cases. Once an integrating factor is determined the integral can be deduced from (\ref{met13}). The $\hat{R}$ and the corresponding integrals are also shown in Table $2$. Our results show that all these integrals are not independent of each other. 

In Table $3$, we have shown all the possible Jacobi last multipliers admitted by Eq.(\ref{exam}), which are obtained from the vector fields $V_1, V_2,...V_8$, using the connection (\ref{delta}). We can now essentially summarize our results on the interconnections between different methods in the form of a pictorial representation as shown in Fig.1.
\begin{figure}[h!]
  \centering
    \includegraphics[width=0.60\textwidth]{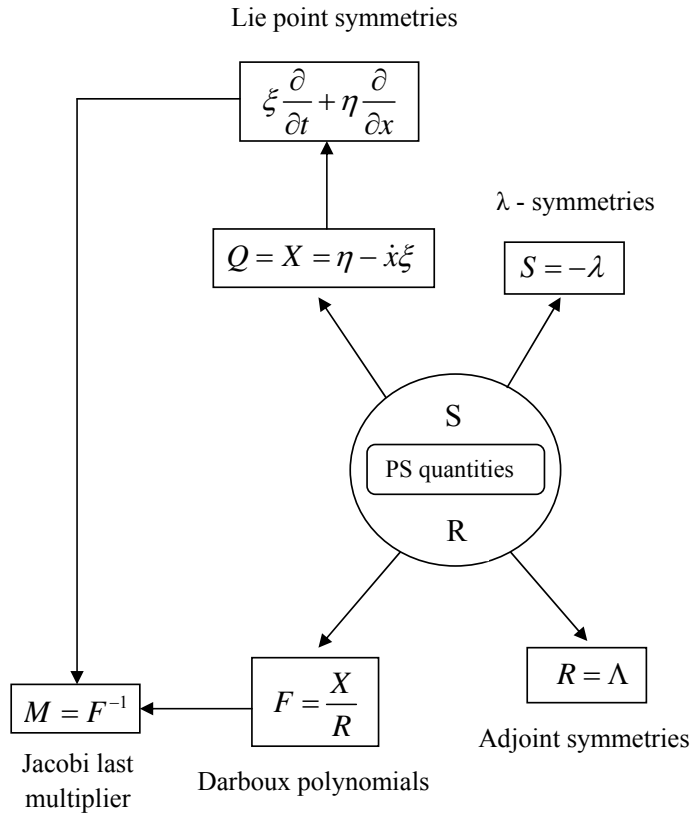}
    \caption{Flow chart connecting Prelle-Singer procedure with other methods}
\end{figure}

\section{Conclusion}
In this paper, we have made a careful analysis of the interconnection between several existing methods to solve second order nonlinear ODEs. For this purpose, we have started with the extended Prelle-Singer method. Two quantities, namely the null form ($S$) and the integrating factor ($R$), play an important role in the Prelle-Singer method. From these quantities, we have brought out the interconnections between the Prelle-Singer method with the several well known methods like Lie symmetries, $\lambda-$symmetries, Darboux polynomials, Jacobi last multiplier and adjoint symmetries methods. Once we know the integrating factor $R$ and null function $S$ from the Prelle-Singer procedure, we are able to derive the Lie symmetries, $\lambda-$symmetries, Darboux polynomials, Jacobi last multiplier, Lagrangian, adjoint symmetries, first integrals and the general solution of a given second order nonlinear ODE. By introducing a suitable transformation in the $S$ equation in the Prelle-Singer method, we identified a connection between the Lie point symmetries and $\lambda-$symmetries methods. By introducing another transformation for the $R$ equation in the Prelle-Singer method we have given the connections between Darboux polynomials, Jacobi last multiplier and adjoint symmetries.  We have demonstrated our assertions with a specific example, namely modified Emden equation. Now we are trying to extend these interconnections to third order ODEs and also to higher order ODEs. The results will be published elsewhere. We believe that the intrinsic connections between different methods shown in this paper will lay foundations to progress further in this area of research.\\\\

RMS acknowledges the University Grants Commission (UGC-RFSMS), Government of India, for providing a Research Fellowship. The work of MS forms part of a research project sponsored by Department of Science and Technology, Government of India.  The work of VKC and ML is supported by a Department of Science and Technology (DST), Government of India, IRHPA research project. ML is also supported by a DAE Raja Ramanna Fellowship and a DST Ramanna Fellowship program.

\begin{landscape}
\begin{figure}[h!]
  \centering
      \includegraphics[width=1.80\textwidth]{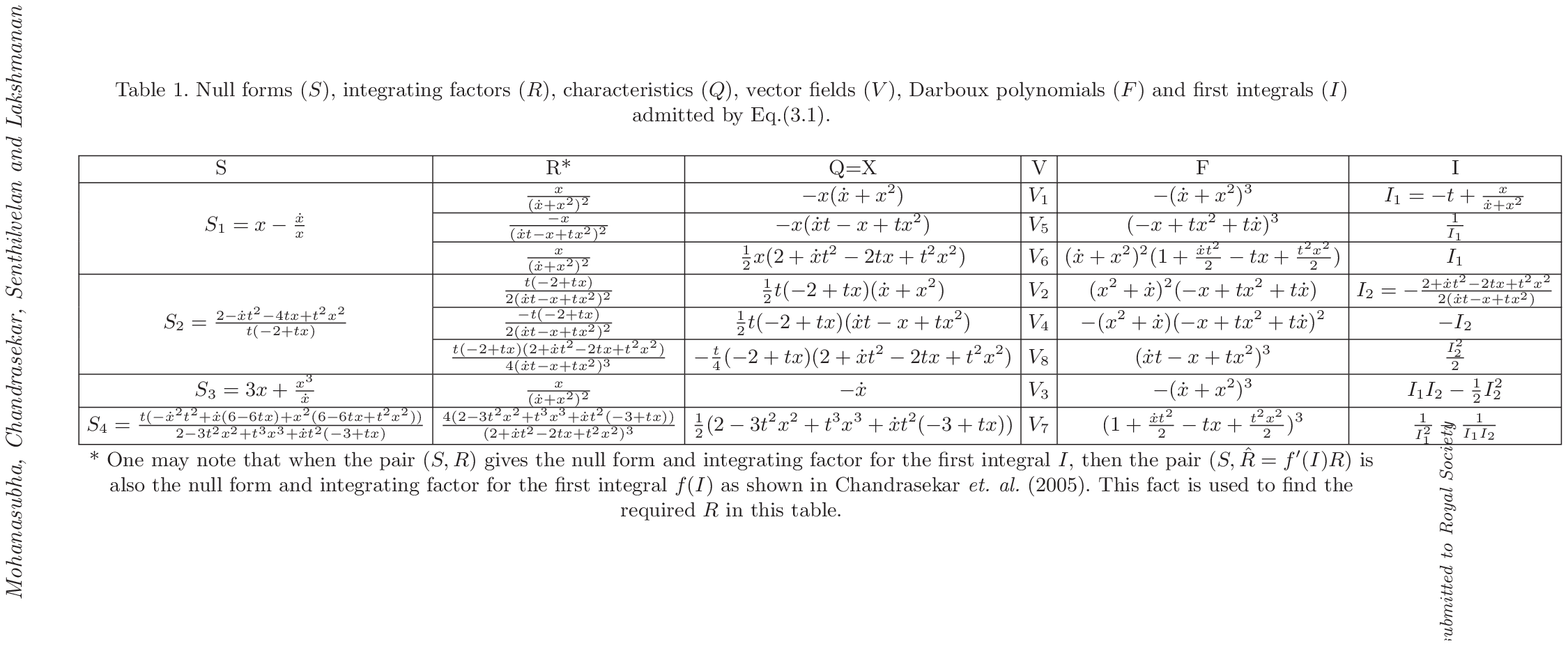}
   \end{figure}
\end{landscape}

\begin{landscape}
\begin{figure}[h!]
  \centering
      \includegraphics[width=1.40\textwidth]{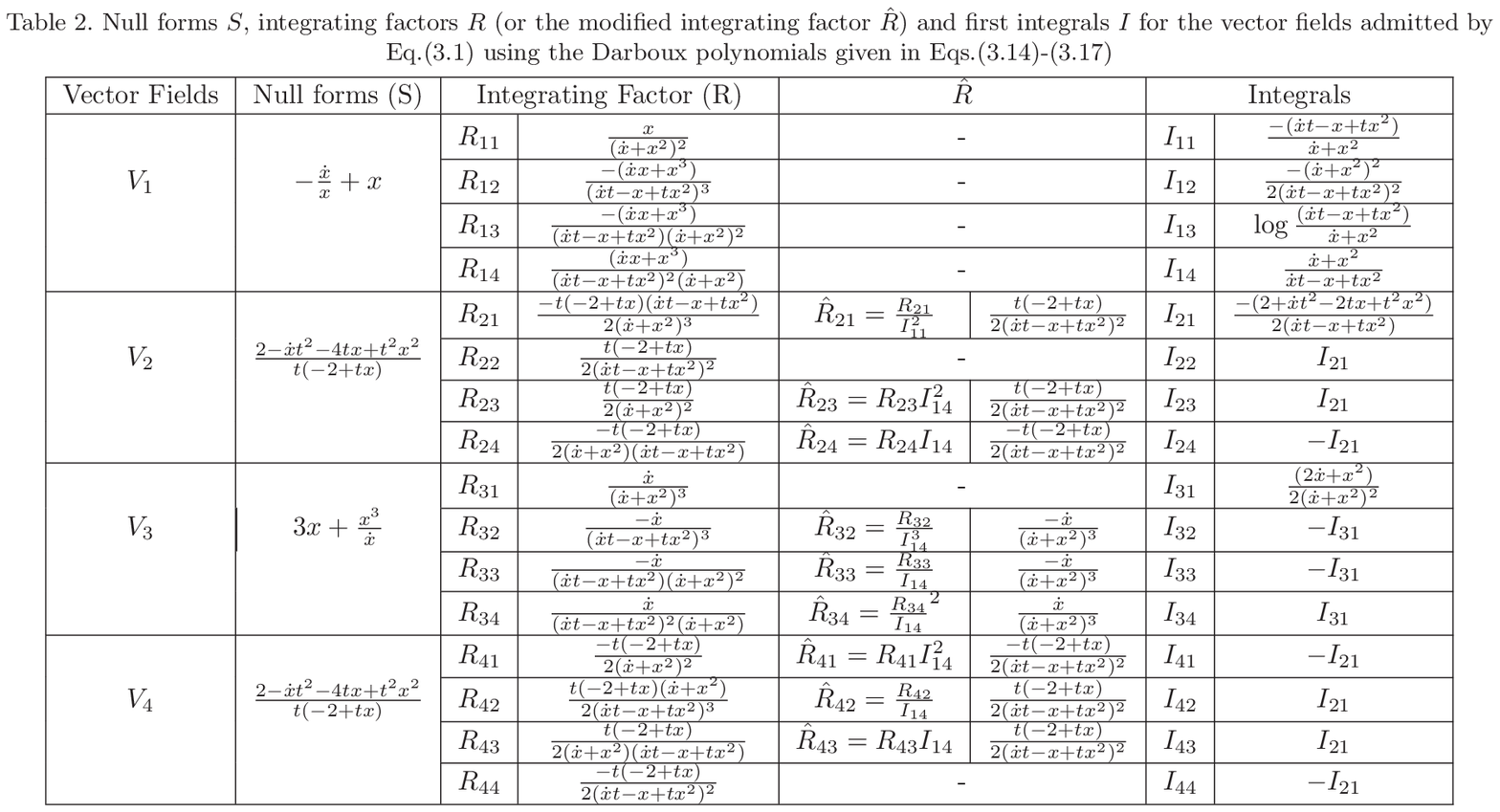}
   \end{figure}
\end{landscape}

\begin{landscape}
\begin{figure}[h!]
  \centering
      \includegraphics[width=1.70\textwidth]{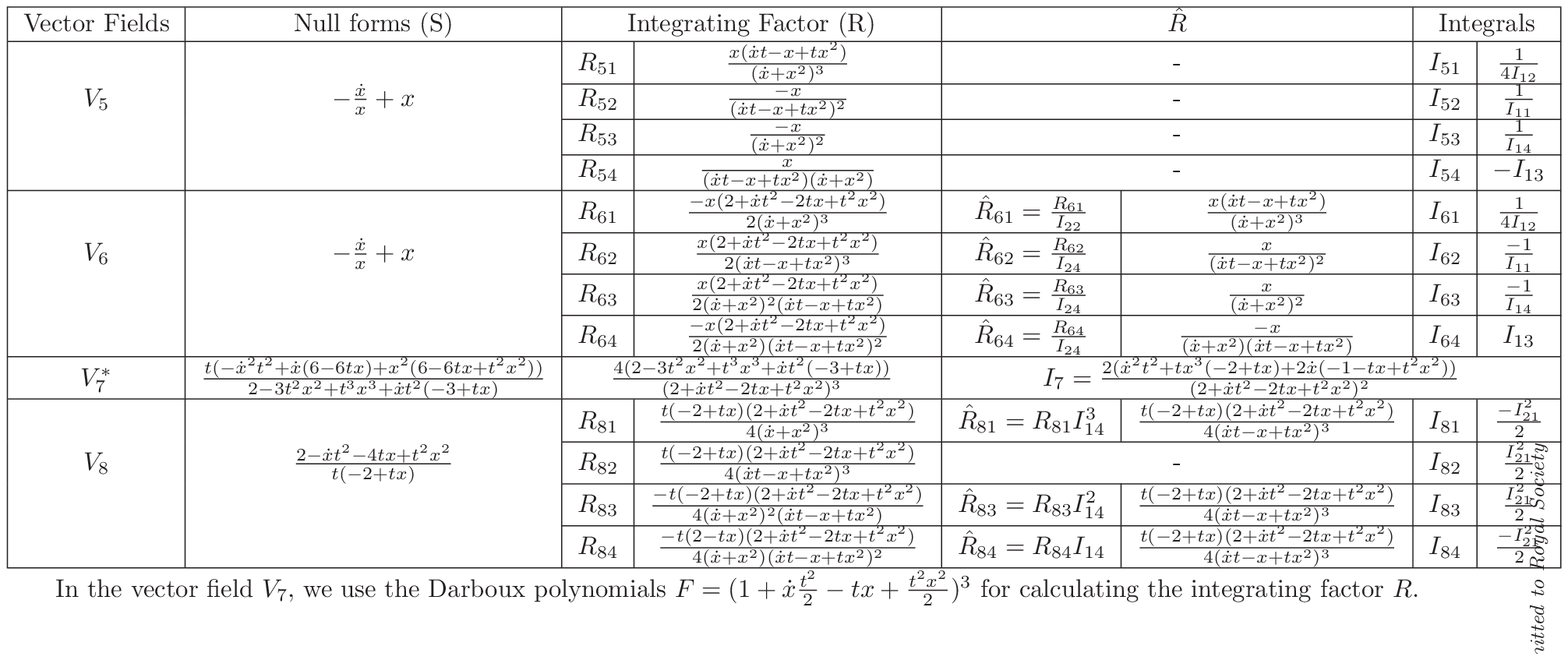}
   \end{figure}
\end{landscape}

\begin{landscape}
\begin{figure}[h!]
  \centering
      \includegraphics[width=1.40\textwidth]{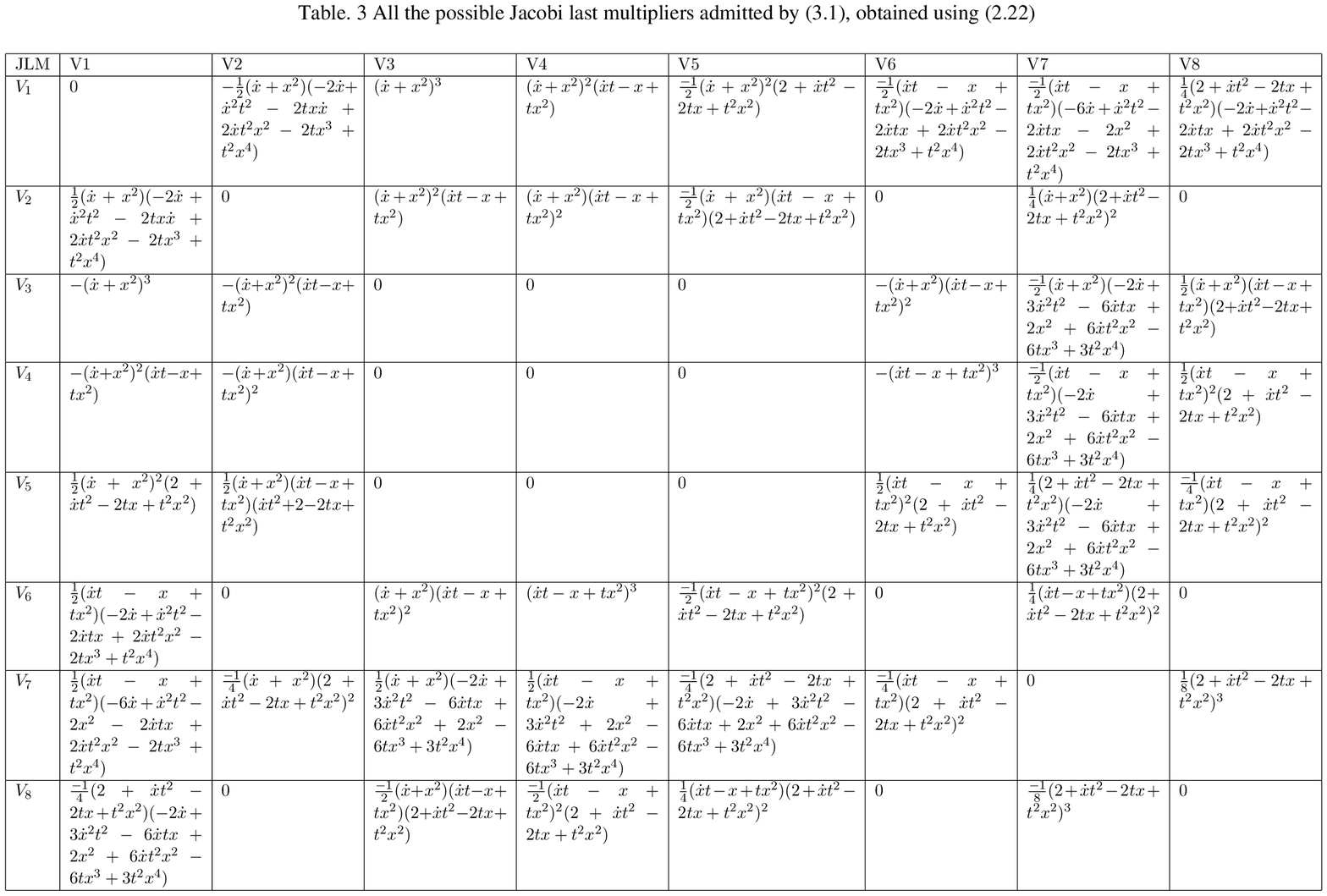}
   \end{figure}
\end{landscape}
\section{Bibiliography}

\end{document}